\documentclass[aps,pra,twocolumn,showpacs]{revtex4-1}
\usepackage{graphics}
\usepackage{graphicx}
\ifx\pdftexversion\undefined
\usepackage[dvips]{hyperref}
\else
\usepackage{hyperref}
\fi
\hypersetup{
  colorlinks = true, linkcolor = blue
}
\begin{document}

\title{Two-body collisions in the time-of-flight dynamics of lattice Bose superfluids}

\author{Antoine Tenart, C\'ecile Carcy, Hugo Cayla, Thomas Bourdel, Marco Mancini, and David Cl\'ement}

\email[Corresponding author: ]{david.clement@institutoptique.fr}

\affiliation{Laboratoire Charles Fabry, Institut d'Optique Graduate School, CNRS, Universit\'e Paris Saclay, 2 Avenue Augustin Fresnel, 91127 Palaiseau cedex, France}

\begin{abstract}
We investigate two-body collisions occurring during the time-of-flight expansion of interacting three-dimensional lattice Bose superfluids. The number of collisions is extracted from the observed {\it s}-wave scattering halos located between the diffraction peaks of the superfluids. These faint halos can be monitored thanks to the large dynamical range in densities associated with detecting individual metastable Helium atoms. We monitor the number of collisions as a function of the total atom number and of the amplitude of the lattice, in a regime where the number of trapped atoms per lattice site is large. In addition, we introduce a classical model of collisions that quantitatively describes the experiment without adjustable parameters. Finally, the present work validates quantitatively the assumption of a ballistic expansion when investigating the Bose-Hubbard Hamiltonian with a unity occupation of the lattice.  
\end{abstract}

% insert suggested PACS numbers in braces on next line
%\pacs{37.10.De, 32.80.Pj, 37.10.Gh, 05.30.Jp}
% insert suggested keywords - APS authors don't need to do this
\keywords{}
%\maketitle must follow title, authors, abstract, \pacs, and \keywords
\maketitle 
 
Quantum gases offer the possibility to investigate momentum distributions in time-of-flight (TOF) experiments \cite{ketterle99}. When considering a single particle, the idea behind this approach is simple: after a sufficiently long ballistic expansion the position of the particle reflects its initial momentum. But when considering an ensemble of interacting atoms, the expansion may be affected by interactions and the ballistic relation questioned. As TOF experiments are extensively used to access information about the momentum-space, a quantitative study of the fidelity with which the in-trap momentum distribution can be obtained is of primary importance.

The role of interaction in TOF experiments has been widely investigated within mean-field approximations. When the trapping frequency along one axis exceeds the mean-field chemical potential, the in-trap momentum distribution is retrieved after a sufficiently long expansion. This includes highly anisotropic traps used to emulate low-dimensional gases \cite{bloch2008} as well as optical lattices \cite{gerbier2008, kupferschmidt2010} but it is not valid for a three-dimensional (3D) harmonically-trapped Bose-Einstein condensate (BEC) \cite{kagan1996, castin1996}. Beyond the mean-field level, two-body collisions must be accounted for, but quantitative investigations have been much more elusive \cite{greiner2001, murthy2014}. Here, we directly measure the number of two-body collisions in the TOF expansion of lattice bosons.

\begin{figure*}[ht!]
\includegraphics[width=2\columnwidth]{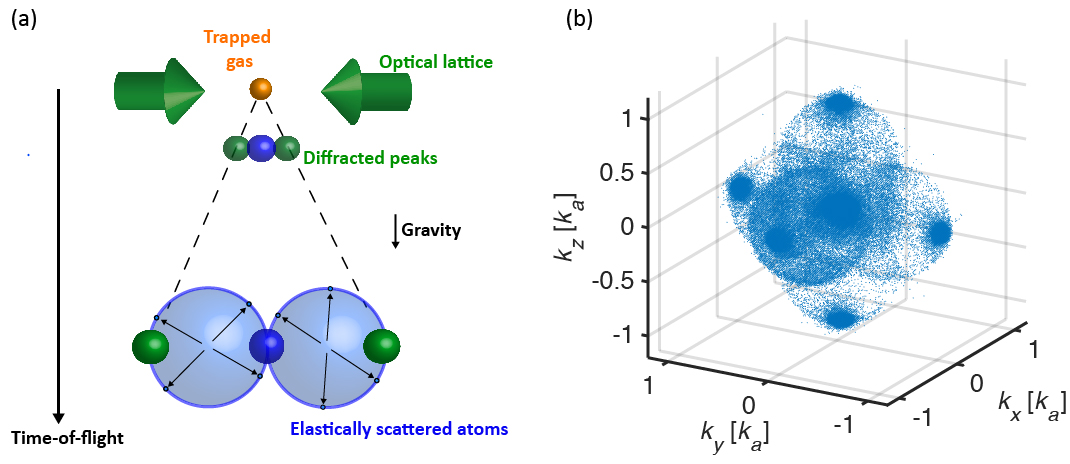}
\caption{ {\bf (a)} Sketch of the experiment along one of the spatial axis (the other two spatial directions are not shown). A Bose-Einstein condensate, initially trapped in an optical lattice, expands in free space after switching off the lattice beams. During the time-of-flight, various copies of the BEC form as a consequence of the diffraction on the in-trap lattice. Elastic two-body collisions between two copies manifest themselves by the presence of spherical halos of scattered atoms. {\bf (b)} Three-dimensional atom distributions of lattice superfluids measured with the single-atom-resolved He$^*$ detector after a $298$ ms time-of-flight. The scattering halos associated to the presence of two-body collisions are clearly visible.}
\label{Fig1}
\end{figure*}

In the ultra-cold regime, {\it s}-wave elastic collisions between two bosons result in the presence of spherical scattering halos defined by momentum and energy conservation. These scattering halos were observed in the collision of Bose-Einstein condensates \cite{kozuma1999, chikkatur2000, thomas2004, katz2005, burdick2016} and are now used as a source of correlated momentum pairs \cite{perrin2007, hodgman2017}. As expected in a classical picture, these experiments show that the number of collisions increases with the atomic density and with the relative velocity of the colliding particles \cite{chikkatur2000}. The TOF distributions of lattice superfluids contain dense components, the diffraction peaks, moving with large relative velocities set by the lattice recoil and two-body collisions are thus expected to take place (see Fig.~\ref{Fig1}(a)). Scattering halos were indeed observed in the expansion from 1D and 2D optical lattices with elongated geometries enhancing the number of collisions \cite{greiner2001, khakimov2016}. To our knowledge a similar observation was not reported in 3D optical lattices with a spherical symmetry and two-body collisions were assumed negligible in landmark experiments \cite{greiner2002, xu2006, trotzky2010, struck2011}.

In this work, we observe the $s$-wave scattering halos present in the TOF distribution of interacting bosons released from a 3D optical lattice and we quantify the number of two-body collisions. For the scattering halos to be visible, we perform experiments with a large lattice filling and we take advantage of the large dynamical range associated with the detection of metastable Helium-4 atoms \cite{cayla2018}. We monitor the number of collisions as a function of the atom number and as a function of the amplitude of the lattice. In addition, we introduce a classical model of collisions that quantitatively reproduces the measured values without any adjustable parameters, in the regime where mean-field interaction is negligible in the TOF dynamics. From extrapolating the results of the model to the regime of unity occupation of the lattice, we establish quantitatively the accuracy with which TOF experiments yield the momentum distribution of lattice bosons. 

In the experiment, we produce Bose-Einstein condensates of metastable Helium-4 ($^4$He$^*$) atoms  \cite{bouton2015} and load them into a 3D optical lattice \cite{cayla2018}. The lattice spacing is $a =\lambda/2=775$~nm where $\lambda$ is the lattice wavelength and its amplitude is denoted $V=s E_{R}$ where $E_{R}=\hbar^2 k_a^2/8m$ is the recoil energy and $k_{a}=2 \pi/a$. We probe the gas with the He$^*$ detector after a TOF $t_{\rm TOF}=298~$ms from which we reconstruct the 3D position $\vec{R}$ of individual atoms \cite{nogrette2015}, in the frame of the center-of-mass of the gas. Recently, we have shown that our apparatus is ideally suited to probe lattice superfluids in the far-field regime, yielding asymptotic momentum densities $n_{\infty}(\vec{k})=(\hbar t_{\rm TOF}/m)^3 \times n_{\rm TOF}(\vec{R}, t_{\rm TOF})$ over several decades \cite{cayla2018}, with $\hbar \vec{k}=m \vec{R}/ t_{\rm TOF}$ and $n_{\rm TOF}(\vec{R}, t_{\rm TOF})$ the density after a time-of-flight $t_{\rm TOF}$. In the present work, we are looking for deviations to the ballistic approximation, in particular for the presence of two-body collisions, with the consequence that $n_{\infty}(\vec{k})\neq n(\vec{k})$, where $n(\vec{k})$ is the in-trap momentum density. To maximise the number of two-body collisions, we load the 3D lattice with a large filling factor, comprised between 3 and 7 atoms per lattice site at the center of the trap. This situation is drastically different from that of our previous works \cite{cayla2018, carcy2019} where we investigated unity-filled lattices. It is obtained by loading BECs with up to $N_{\rm bec} \sim 6 \times 10^5$ atoms in the 3D lattice. In addition, we use lattice amplitudes larger than $s \gtrsim4$ to ensure that the short-time expansion is ballistic, {\it i.e.} that the zero-point energy of a lattice site largely exceeds the chemical potential of the gas. 

An example of a 3D single-atom-resolved distribution is shown in Fig.~\ref{Fig1}(b). We clearly distinguish the presence of spherical scattering halos between the diffraction peaks at $| \vec{k} |=k_{a}$ and the central peak $| \vec{k} |=0$. These halos are an unambiguous signature of the presence of elastic two-body collisions. The atomic density of the halos is much smaller than that of the diffraction peaks (by a factor $\sim10^{-7}$) because the fraction of  atoms which collide is small and because the volume over which these scattered atoms distribute is much larger than that of a BEC peak, by a factor $(L/a)^2\gg1$ where $L\gtrsim 30 a$ is the Thomas-Fermi radius of the trapped gas \cite{NoteRatioVolume}. This implies that revealing the scattering halos requires a large dynamic range in density, as that provided by the He$^*$ detector \cite{nogrette2015, chang2016}.

We now introduce a model to estimate the number of two-body collisions taking place during the TOF expansion of a lattice superfluid. To this aim, we elaborate a classical model inspired by the one introduced in the context of two colliding BECs \cite{band2000, zin2005, zin2006}. It was shown that such a classical model of binary collisions is in excellent agreement with a quantum model to first order in perturbation theory when the sound velocity of the BECs is smaller than the relative velocity between the two clouds \cite{zin2006}. We also note that a classical model does not include bosonic stimulation but this effect is expected to be negligible when the number of scattering particles per mode is much smaller than unity. This two hypotheses are valid in the regime of our experiment, as it will be discussed below.

We start by considering the lattice Bose-Einstein Condensate in the trap with $N_{\rm bec}$ atoms. In the following, we neglect the role of depleted atoms whose atomic density is too small to contribute substantially to the number of collisions. In the regime of large lattice filling, the average density $n_{\rm bec}(\vec{r})$ in the trap (or similarly the atom number distribution in the lattice) is well approximated by the parabolic profile associated with the Thomas-Fermi approximation. The in-trap Thomas-Fermi radius $L$ can be obtained from the atom number $N_{\rm bec}$ and the trapping frequency $\omega/2\pi \sim 140 \sqrt{s}$~Hz using an interaction strength which accounts for the lattice potential \cite{kramer2002}. The distribution of quasi-momenta $\vec{q}$ of the trapped BEC is centered on $\vec{q}=\vec{0}$ with a typical RMS width $\Delta q \sim 1.58/L$ \cite{stenger1999}. When the lattice potential is abruptly switched off at time $t=0$, several copies of the BEC are created along each axis of the 3D lattice whose average momentum equals to $ j \times k_{a}$ where $j \in \mathbf{Z}$. On the one hand, at the time the lattice is switched off, the densities of the BEC copies are identical, up to a scaling factor $\alpha_{j}$ setting the number of atoms in the copy labelled $j$, $N_{j}=\alpha_{j} N_{\rm bec}$. The coefficients $\alpha_{j}$ depend on the amplitude of the lattice $s$ and are calculated numerically from the Fourier transform of the Wannier function associated with a lattice site \cite{ashcroft1976}. On the other hand, the momentum width $\Delta k$ of the different copies is identical to that of the trapped BEC, $\Delta k=\Delta q$. Any two copies have a large relative velocity (of the order of the recoil velocity $v_{a}=\hbar k_{a}/m$) and atoms from the two copies have a non-negligible probability to collide. In the following, we concentrate the discussion of two specific copies: the central diffraction peak $j=0$ and the first-order peak $j=1$ moving along the $x$-axis (see Fig.~\ref{Fig1}(a)). By symmetry the number of collisions between the central peak and any other first-order peak is identical to the one we consider. In addition, the collisions between other copies of the BEC, say for instance between two first-order peaks or between the central peak and a second-order peak, are hardly visible in the experimental data. We will yet discuss their contribution towards the end of the article.

To calculate the number of collisions between the copies $j=0$ and $j=+1$, we evaluate the collision rate of one atom located at position $\vec{r}$ at time $t$ in the BEC copy $j=0$ with all the atoms of the copy $j=1$. This rate writes $\Gamma_{\rm coll} = n_{1}(\vec{r},t) \times \sigma \times v_{0,1}$ where $n_{1}$ is the density of the copy $j=1$, $v_{0,1}=v_{a}$ is the relative velocity of the two copies and $\sigma=8 \pi a_{s}^2$ is the scattering cross-section, with $a_{s}$ the s-wave scattering length ($a_{s}\simeq142 a_{0}$ for $^4$He$^*$ atoms, with $a_{0}$ the Bohr radius). The total number of collisions between the two copies results from the integration over all the particles of the copy $j=0$ and over the time interval during which the copies spatially overlap. It can be multiplied by 6 to get the total number of collisions with all first-order copies:

\begin{equation}
N_{\rm coll}= 6 \nonumber  \int dt \int d\vec{r} \  \sigma \ v_{a} \  n_{0}(\vec{r},t) \ n_{1}\left(\vec{r},t \right). 
\label{Eq:N_coll_integral}
\end{equation}

To proceed with the evaluation of Eq.~\ref{Eq:N_coll_integral}, the TOF dynamics of the densities $n_{j}(\vec{r},t)$ must be known. A full description of the TOF dynamics is beyond the scope of this article. Instead we provide a simpler physical description of the TOF relying on the different energy and time scales of the problem. The shortest time scale is set by the frequency of a lattice site $\omega_{\rm site}$, and the corresponding harmonic oscillator length $a_{\rm h.o.}=\sqrt{\hbar/m \omega_{\rm site}}$. On the time scale $t_{\rm 0} = m a a_{\rm h.o.}/ h \sim 1$~$\mu$s, the wave functions of separated lattice sites overlap. After a few $t_{0}$, the density profile of the copies is smoothen, with a lower density than that in the trap and a total size of the gas that is hardly larger than the in-trap size $L$. The density profiles of the copies are thus well described by the parabolic profile $n_{\rm bec}(\vec{r})$. 

The other time scales in the problem are much larger than $t_0$ and associated with {\it (i)} the spatial separation of the two copies, $t_{\rm sep} \sim 2 L/v_{a} \sim 0.1$~ms; {\it (ii)} the expansion of a BEC copy driven by its kinetic energy, $t_{\rm kin} \sim m L / \hbar \Delta k \sim 10$~ms ; and {\it (iii)} the expansion of a BEC copy under the effect of the mean-field interacting potential, $t_{\rm MF}$. For trapped cloud sizes much larger than the lattice spacing, $L\gg a$, one has $\Delta k \ll k_{a}$ and, in turn, $t_{\rm sep}\ll t_{\rm kin}$. On the contrary, $t_{\rm MF}$ depends on the atom number and it can be comparable to $t_{\rm sep}$, paving the way to two possible scenarios. When the atom number is small enough, the effect of the mean-field potential is negligible and the density profile of the copies does not change while they separate spatially. When it cannot be neglected, the cloud sizes are enlarged before the copies separate as a result of the repulsive mean-field potential. 

We first consider the scenario where the effect of mean-field potential is negligible. The density profile of each copy is constant over the duration $t_{\rm sep}$ since $t_{\rm sep}\ll t_{\rm kin}$. Evaluating the expression of Eq.~\ref{Eq:N_coll_integral} with a parabolic profile for the BEC density of radius $L$, we obtain%
\begin{equation}
N_{\rm coll}= \frac{48 \alpha_{0} \alpha_{1}}{315}   \left (\frac{15 N_{\rm bec}  a_{s}}{L} \right )^2.  \label{Eq:Ncoll}
\end{equation}
The scaling $N_{\rm coll} \propto a_{s}^2 N_{\rm bec}^2/L^2$ is identical to that found previously in the description of the collisions between two BECs \cite{zin2005,zin2006}. On the one hand, the ratio $(N_{\rm bec}/L)^2$ sets the variation of $N_{\rm coll}$ with the total atom number (see Eq.~\ref{Eq:Ncoll}), yielding in the Thomas-Fermi approximation $N_{\rm coll} \propto N_{\rm bec}^{8/5}$. On the other hand, the lattice amplitude enters both in the product $\alpha_{j}\alpha_{j'}$ and in the size $L$.  As $L$ only slightly decreases with the lattice amplitude, the main variation of $N_{\rm coll}$ is thus driven by the product $\alpha_{0}\alpha_{1}$. This product increases with $s$ at small lattice amplitude as a larger number of atoms populates the first order diffraction peak $j=1$. At larger lattice amplitude, $\alpha_{0}\alpha_{1}$ decreases as the total population in the $j=0$ and $j=1$ peaks decreases to populate diffraction peaks of higher orders. 
 
When the effect of the mean-field potential is not negligible, the expansion of the copies is faster than in the ballistic case that we have just described. In this scenario the interaction-induced expansion decreases the density of the copies and, in turn, the number of collisions as compared to that predicted by Eq.~\ref{Eq:Ncoll}. Interestingly, the momentum width of the scattering halos $\delta k_{\rm s}$ after a long TOF provides a direct mean to identify when the mean-field potential plays a non-negligible role. When the expansion is ballistic, $\delta k_{\rm s}$ is equal to the in-trap momentum width $\Delta k \propto 1/L$ and  decreases with the total atom number $N_{\rm bec}$ as $L$ increases with $N_{\rm bec}$. On the contrary, when the expansion is affected by mean-field interactions, $\delta k_{\rm s}$ reflects the increased kinetic energy associated with the mean-field interaction potential and thus increases with $N_{\rm bec}$. As a result, the width $\delta k_{\rm s}$ is expected to take a minimum value as a function of $N_{\rm bec}$ that identifies the crossover between the two scenarios. 

\begin{figure}[h!]
\includegraphics[width=\columnwidth]{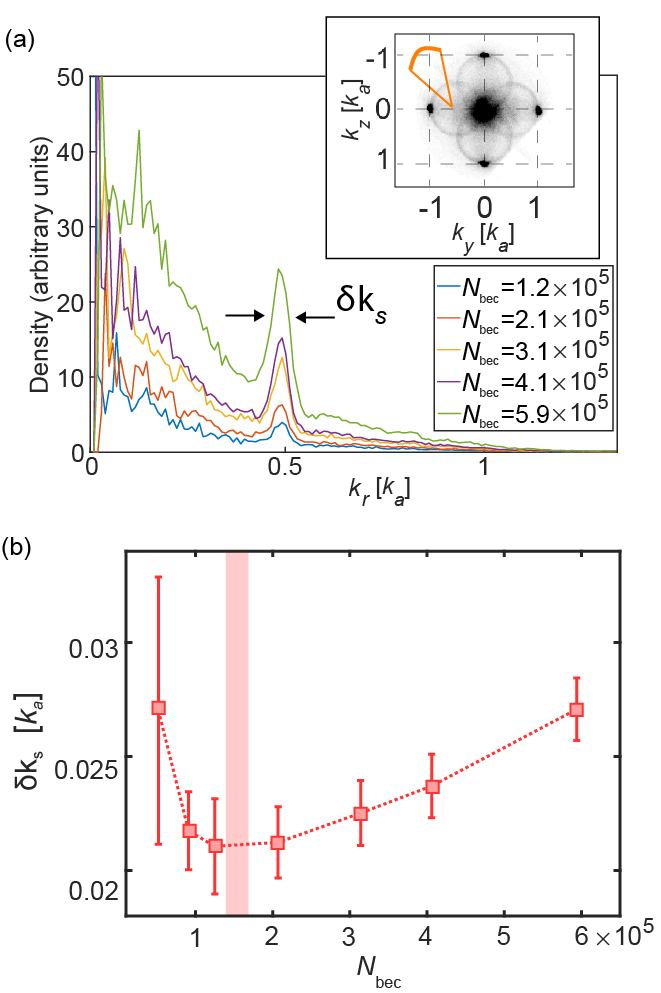}
\caption{{\bf (a)} Atom number histograms plotted as a function of the distance $k_{r}$ to the center of the halo. Histograms at a fixed amplitude $s=5$ and for various atom numbers $N_{\rm bec}$ (from $N_{\rm bec}=125 \times 10^3$ to $N_{\rm bec}=594 \times 10^3$) are shown. Inset: Two-dimensional cut at $k_{x}=0$ through the 3D distribution. The region where the number of scattered atoms is calculated is delimited by the orange line. Note that to make the scattering halos visible, the diffraction peaks are highly saturated (by a factor about $\sim 30$) {\bf (b)} RMS width $\delta k_{\rm s}$ of the scattering halo as function of $N_{\rm bec}$. The red dashed line is a guide to-the-eye. The red shaded area signals the crossover  between the two scenarios exposed in the main text.}
\label{Fig2}
\end{figure}

In the experiment, we measure the number of collisions as follows. For each halo, we restrict the analysis to a small slice that excludes the volumes where the halo intersects with the condensate peaks or with other halos, as illustrated in the inset of Fig.~\ref{Fig2}(a). We then calculate the distance $k_r$ between an atom and the center of the scattering halo, and build a histogram $\mathcal{N}(k_r)$ of the atom number located at a distance $k_r$ (the bin size of the histogram is $0.01 k_a$). In doing so, we also take into account the detection efficiency $\alpha_{D}$ by multiplying the atom number in each bin of the histogram by $1/\alpha_{D}$. We have carefully calibrated $1/\alpha_{D}\simeq 15.5(1.0)$  by recording the diluted distributions of Mott insulators, in order to avoid the saturation effects associated with the He$^*$ detector \cite{nogrette2015}. Examples of histograms are plotted in Fig.~\ref{Fig2}(a) for various total atom numbers and at a fixed lattice amplitude $s=5$. The scattering halo can be identified as the peak located at $k_r=0.50(1)k_{a}$. After removing the contribution from the background density (due to the quantum and the thermal depletion \cite{chang2016}), we fit the peak with a Gaussian function to extract the width $\delta k_{s}$ and the number of scattered atoms in the investigated slice. Assuming a spherical symmetry, the total number of scattered atoms in one halo is obtained by integrating the value measured in the slice over the entire scattering sphere. Finally, the measured number of collisions $N_{\rm coll}^{\rm exp}$ between the copies $j=0$ and $j=1$ is obtained from summing the number of scattered atoms measured in the various observed scattering halos \cite{NoteScattHalo}. 

The measured RMS momentum width $\delta k_{s}$ of the scattering halo is plotted in Fig.~\ref{Fig2}(b) as a function of the atom number $N_{\rm bec}$. We observe that $\delta k_{s}$ has a minimum value which signals the crossover between the two scenarios introduced previously.  From Fig.~\ref{Fig2}(b), we conclude that for atom numbers larger than $N_0 \simeq 1.7 \times 10^5$, the model associated to Eq.~\ref{Eq:Ncoll} should overestimate the number of collisions.

\begin{figure}[h!]
\includegraphics[width=\columnwidth]{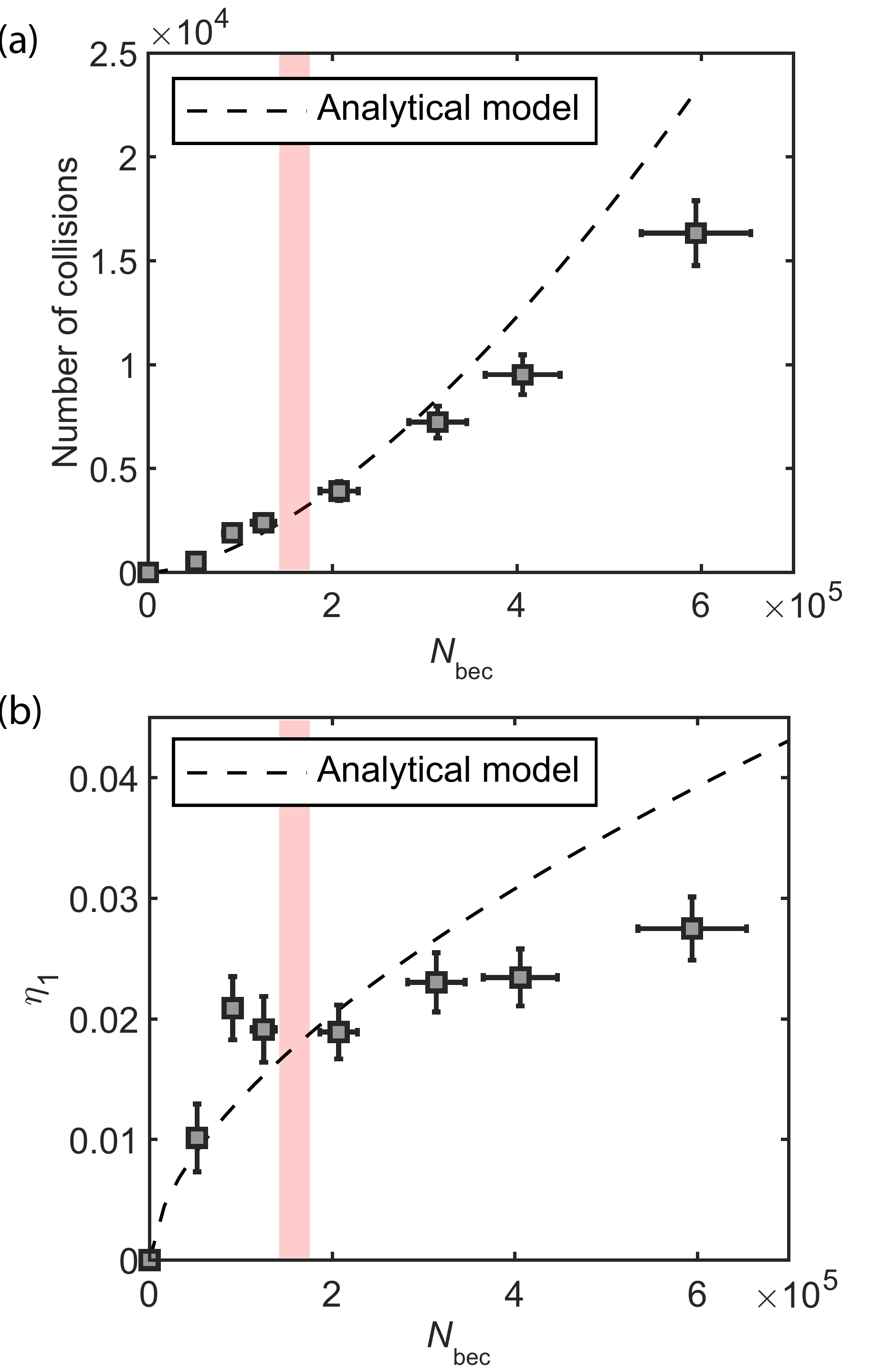}
\caption{(a) Number of collisions  $N_{\rm coll}^{\rm exp}$ between the copies $j=0$ and $j=1$ plotted as a function of the atom number $N_{\rm bec}$, at a fixed lattice amplitude $s=5$. The vertical error bars account for the standard error of the mean on $N_{\rm coll}^{\rm exp}$ and the uncertainty on the detection efficiency. The horizontal error bars depict the standard deviation on $N_{\rm bec}$. The dashed line is the prediction of the model of collision.  (b)  Probability of collision per atom $\eta_{1}=N_{\rm coll}^{\rm exp}/N_{\rm bec}$ as a function of $N_{\rm bec}$.}
\label{Fig3}
\end{figure}

In Fig.~\ref{Fig3}(a) we plot the number of collisions $N_{\rm coll}^{\rm exp}$ as a function of $N_{\rm bec}$ for a fixed amplitude of the lattice $s=5$ along with the prediction of Eq.~\ref{Eq:Ncoll}. We observe that $N_{\rm coll}^{\rm exp}$ increases with $N_{\rm bec}$ faster than linearly, as expected from Eq.~\ref{Eq:Ncoll}. At low atom numbers, we find a good agreement with the prediction of the classical model with no adjustable parameters. The number of collisions becomes lower than the predicted values for atom numbers larger than the value $N_{0}$ identified from the measurement of $\delta k_{s}$ (see Fig.~\ref{Fig2}) in agreement with the physical picture previously described. Note that the number of modes in the scattering halos is of order $6 \times (L/a)^2\sim10^4$ \cite{NoteRatioVolume}, which implies that the population of the modes is smaller than one at low atom numbers. Finally, we plot the probability of collision per atom between the BEC copy $j=0$ and all the copies $j=1$, $\eta_{1}=N_{\rm coll}^{\rm exp}/N_{\rm bec}$, in Fig.~\ref{Fig3}(b). We observe that $\eta_{1}$ is of the order of a few percent in the regime we investigated of large fillings of the lattice.

\begin{figure}[h!]
\includegraphics[width=\columnwidth]{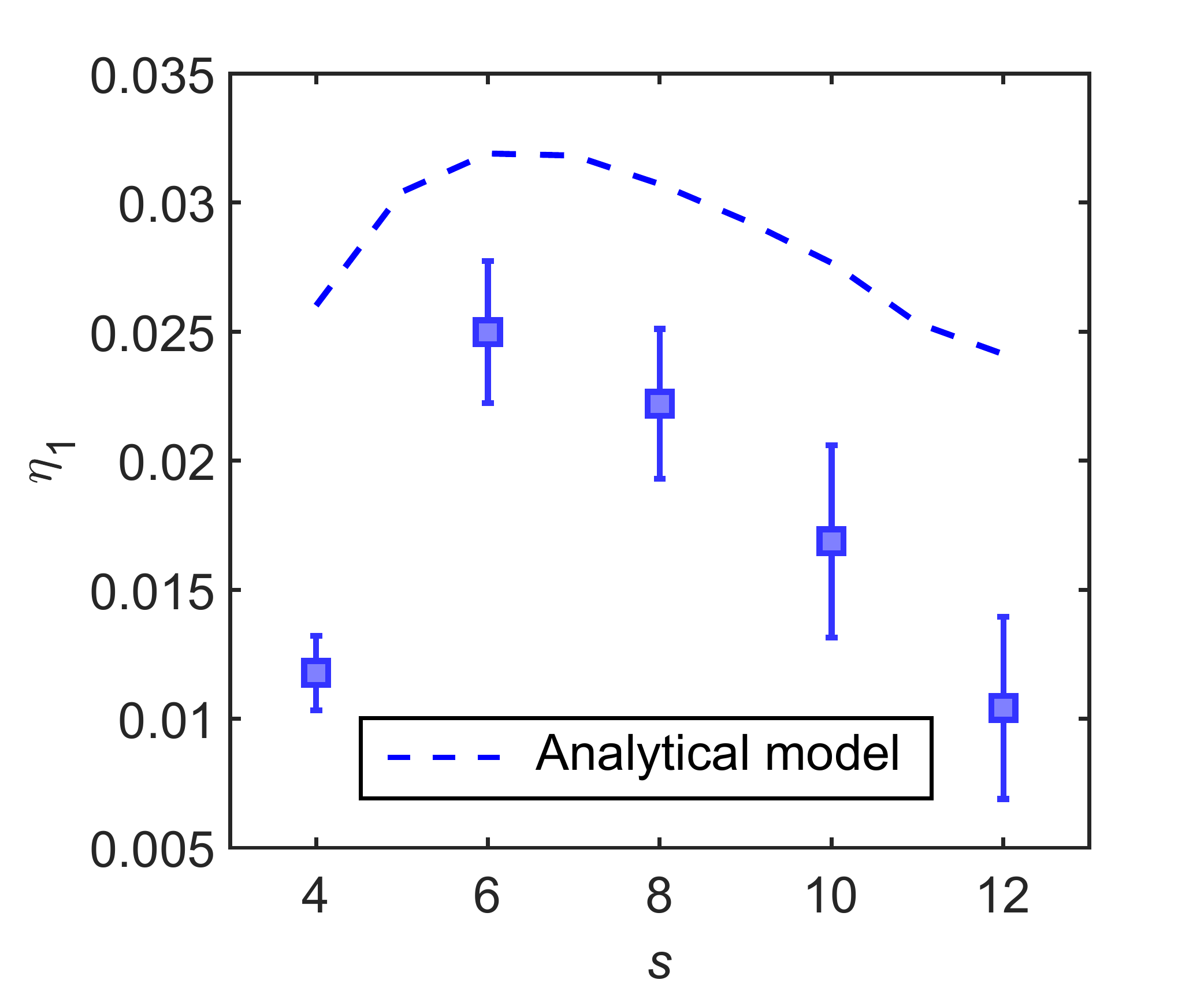}
\caption{Probability of collision per atom $\eta_{1}$ as a function of $s$ at a fixed atom number $N_{\rm bec}=3.9(4) \times 10^5$. The blue dashed line is the prediction of the model of collision.}
\label{Fig4}
\end{figure}

In Fig.~\ref{Fig4} we plot $\eta_1$ as a function of the lattice amplitude $s$ for a fixed atom number $N=3.9(4)\times 10^5$. The number of collisions is found to increase with $s$ at small amplitudes while it decreases with $s$ at $s>5$. This non-monotonic behavior is well reproduced by the classical model of collisions and it can be understood by considering the populations of the different BEC copies, as explained previously. As expected, $\eta_{1}$ is found to be lower than predicted by the classical model since the atom number is larger than $N_0$ where the model over-estimates the number of collisions.

So far we have considered only the atoms scattered in the collision of the copies $j=0$ and $j=1$ because the corresponding scattering halos are clearly visible in the experiment. One also expects that collisions between two copies $j=1$, as well as collisions involving copies corresponding to higher-order of diffraction $j>2$, take place but the associated scattering halos are more diluted (because of their larger volume and of a lower number of collisions) and we do not observe them in the experiment. At increasing amplitude $s$ of the lattice, the collisions that we have not considered so far may contribute substantially to the total number of collisions. To evaluate these contributions, one can use Eq.~\ref{Eq:Ncoll} with the appropriate $\alpha_{j}$'s, accounting for the number of pairs of copies. An upper bound for the total number of collisions is provided by considering an extreme situation where the number of copies would be as large as the number of atoms. This upper bound is only a factor 3 larger than $\eta_{1}$ and as a result the measured probability of collision $\eta_{1}$ also provides a good estimate for the total probability of collisions. Importantly, one can use Eq.~\ref{Eq:Ncoll} to estimate the number of collisions occurring in an experiment with a unity filling of the lattice sites, a situation where we are not able to observe any scattering halos \cite{cayla2018}. We find that the probability for an atom to collide during the TOF is extremely low, $\eta_{1} \sim 10^{-4}$. 

In conclusion, we have quantitatively studied the two-body collisions occurring during the time-of-flight expansion of lattice superfluids with a large lattice filling. We have introduced a classical model of collisions which is found in agreement with the measured number of collisions without adjustable parameters, in the regime where mean-field interactions do not affect the TOF dynamics. From extrapolating the results of the model to the regime of unity lattice filling, we find that the probability for an atom to collide during the TOF is extremely low. In other regions of the finite-temperature phase diagram of the Bose-Hubbard Hamiltonian with a unity filling -- at higher temperature or when approaching the Mott insulating regime -- the probability of collision would be further reduced. From these considerations, the present work confirms the accuracy of the mapping $n_{\infty}(\vec{k}) \simeq n(\vec{k})$ for unity-filled lattices beyond the mean-field level \cite{cayla2018}. It also validates the possibility to investigate the momentum-space correlations between individual atoms in TOF experiments \cite{carcy2019}.

%%%%%%%%%%%%%%%%%%%%%%%%%%%%%%%%%%%%%%%%%%%%%%%%%%%%%%%%%
\vspace{5mm}
\begin{acknowledgments}
We acknowledge fruitful discussions with F. Gerbier and the members of the Quantum Gases group at Institut d'Optique. We thank D. Boiron for a careful reading of the manuscript. This work benefited from financial support by the LabEx PALM (Grant number ANR-10-LABX-0039), the Institut Francilien de Recherche sur les Atomes Froids, the ``Fondation d'entreprise iXcore pour la Recherche" and the Agence Nationale pour la Recherche (Grant number ANR-17-CE30-0020-01). D.C. acknowledges support from the Institut Universitaire de France.
\end{acknowledgments}

% Create the reference section using BibTeX:
%\bibliography{}

\begin{thebibliography}{21}
\expandafter\ifx\csname natexlab\endcsname\relax\def\natexlab#1{#1}\fi
\expandafter\ifx\csname bibnamefont\endcsname\relax
  \def\bibnamefont#1{#1}\fi
\expandafter\ifx\csname bibfnamefont\endcsname\relax
  \def\bibfnamefont#1{#1}\fi
\expandafter\ifx\csname citenamefont\endcsname\relax
  \def\citenamefont#1{#1}\fi
\expandafter\ifx\csname url\endcsname\relax
  \def\url#1{\texttt{#1}}\fi
\expandafter\ifx\csname urlprefix\endcsname\relax\def\urlprefix{URL }\fi
\providecommand{\bibinfo}[2]{#2}
\providecommand{\eprint}[2][]{\url{#2}}

\bibitem{ketterle99} W. Ketterle, D. S. Durfee and D. M. Stamper-Kurn, {\it Making, probing and understanding Bose-Einstein condensates}, Proceedings of the International School of Physics "Enrico Fermi", {\bf 140}, 67-176 (1999)

\bibitem{bloch2008} I. Bloch, J. Dalibard and W. Zwerger, {\it Many-body physics with ultracold gases}, Rev. Mod. Phys. {\bf 80}, 885 (2008).

\bibitem{gerbier2008} F. Gerbier, S. Trotzky, S. Folling, U. Schnorrberger, J. D. Thompson, A. Widera, I. Bloch, L. Pollet, M. Troyer, B. Capogrosso-Sansone, N.V. Prokofev, and B.V. Svistunov, {\it Expansion of a Quantum Gas Released from an Optical Lattice}, Phys. Rev. Lett. {\bf 101}, 155303 (2008).

\bibitem{kupferschmidt2010} J. N. Kupferschmidt and E. J. Mueller, {\it Role of interactions in time-of-flight expansion of atomic clouds from optical lattices}, Phys. Rev. A {\bf 82}, 023618 (2010).

\bibitem{castin1996} Y. Castin and R. Dum, {\it Bose-Einstein condensates in time dependent traps}, Phys. Rev. Lett.  {\bf 77}, 5315 (1996).

\bibitem{kagan1996} Y. Kagan, E. L. Surkov and G. V. Shlyapnikov, {\it Evolution of a Bose-condensed gas under variations of the confining potential}, Phys. Rev. A {\bf 54}, 1753(R) (1996).

\bibitem{greiner2001} M. Greiner, I. Bloch, O. Mandel, T. W. Hansch, and T. Esslinger, {\it Exploring phase coherence in a 2D lattice of Bose-Einstein condensates}, Phys. Rev. Lett. {\bf 87}, 160405 (2001).

\bibitem{murthy2014} P. A. Murthy, D. Kedar, T. Lompe, M. Neidig, M. G. Ries, A. N. Wenz, G. Zurn and S. Jochim, {\it Matter wave Fourier optics with a strongly interacting two-dimensional Fermi gas}, Phys. Rev. A {\bf 90}, 043611 (2014).

\bibitem{kozuma1999} M. Kozuma, L. Deng, E. W. Hagley, J. Wen, R. Lutwak, K. Helmerson, S. L. Rolston, and W. D. Phillips, {\it Coherent Splitting of Bose-Einstein Condensed Atoms with Optically Induced Bragg Diffraction}, Phys. Rev. Lett {\bf 82} 871 (1999).

\bibitem{chikkatur2000} A. P. Chikkatur, A. Gorlitz, D. M. Stamper-Kurn, S. Inouye, S. Gupta, and W. Ketterle, {\it Suppression and Enhancement of Impurity Scattering in a Bose-Einstein Condensate}, Phys. Rev. Lett. {\bf 85}, 483 (2000).

\bibitem{thomas2004} N. R. Thomas, N. Kjaergaard, P. S. Julienne, and A. C. Wilson, {\it Imaging of s and d Partial-Wave Interference in Quantum Scattering of Identical Bosonic Atoms}, Phys. Rev. Lett. {\bf 93}, 173201 (2004).

\bibitem{katz2005} N. Katz, E. Rowen, R. Ozeri, and N. Davidson, {\it Collisional Decay of a Strongly Driven Bose-Einstein Condensate}, Phys. Rev. Lett. {\bf 95}, 220403 (2005).

\bibitem{burdick2016} N. Q. Burdick, A. G Sykes, Y. Tang and B. L Lev, {\it Anisotropic collisions of dipolar Bose–Einstein condensates in the universal regime}, New Journal of Physics, {\bf 18} 111004 (2016).

\bibitem{perrin2007} A. Perrin, H. Chang, V. Krachmalnicoff, M. Schellekens, D. Boiron, A. Aspect, and C. I. Westbrook, {\it Observation of Atom Pairs in Spontaneous Four-Wave Mixing of Two Colliding Bose-Einstein Condensates}, Phys. Rev. Lett. {\bf 99}, 150405 (2007).

\bibitem{hodgman2017} S. S. Hodgman, R. I. Khakimov, R. J. Lewis-Swan, A. G. Truscott and K. V. Kheruntsyan, {\it Solving the Quantum Many-Body Problem via Correlations Measured with a Momentum Microscope}, Phys. Rev. Lett. {\bf 118}, 240402 (2017).

\bibitem{khakimov2016} R. I. Khakimov, B. M. Henson, D. K. Shin, S. S. Hodgman, R. G. Dall, K. G. H. Baldwin and A. G. Truscott, {\it Ghost imaging with atoms}, Nature {\bf 540}, 100-103 (2016).

\bibitem{greiner2002} M. Greiner,  O. Mandel, T. Esslinger, T. W. Hansch and I. Bloch, {\it Quantum phase transition from a superfluid to a Mott insulator in a gas of ultracold atoms}, Nature {\bf 415}, 39-44 (2002).

\bibitem{xu2006} K. Xu, Y. Liu, D. E. Miller, J. K. Chin, W. Setiawan and W. Ketterle, {\it Observation of Strong Quantum Depletion in a Gaseous Bose-Einstein Condensate}, Phys. Rev. Lett., {\bf96,} 180405 (2006).

\bibitem{struck2011} J. Struck, C. Olschlager, R. Le Targat, P. Soltan-Panahi, A. Eckardt, M. Lewenstein, P. Windpassinger and K. Sengstock, {\it Quantum Simulation of Frustrated Classical Magnetism in Triangular Optical Lattices}, Science {\bf 333}, 996--999 (2011).

\bibitem{trotzky2010} S. Trotzky, L. Pollet, F. Gerbier, U. Schnorrberger, I. Bloch, N. V. Prokof’ev, B. Svistunov and M. Troyer, {\it Suppression of the critical temperature for superfluidity near the Mott transition}, Nature Physics {\bf 6}, 998-1004 (2010).

\bibitem{cayla2018} H. Cayla, C. Carcy, Q. Bouton, R. Chang, G. Carleo, M. Mancini and D. Cl\'ement, {\it Single-atom-resolved probing of lattice gases in momentum space}, Phys. Rev. A Rapid Comm. {\bf 97} 061609 (2018).

\bibitem{bouton2015} Q. Bouton, R. Chang, A. L. Hoendervanger, F. Nogrette, A. Aspect, C. I. Westbrook and D. Cl\'ement, {\it Fast production of Bose-Einstein condensates of metastable Helium}, Phys. Rev. A {\bf 91}, 061402(R) (2015).

\bibitem{nogrette2015}  F. Nogrette, D. Heurteau, R. Chang, Q. Bouton, C. I. Westbrook, R. Sellem and D. Cl\'ement, {\it Characterization of a detector chain using a FPGA-based Time-to-Digital Converter to reconstruct the three-dimensional coordinates of single particles at high flux}, Rev. Scient. Intrum. {\bf 86}, 113105 (2015).

\bibitem{carcy2019} C. Carcy, H. Cayla, A. Tenart, A. Aspect, M. Mancini and D. Cl\'ement, {\it Momentum-space atom correlations in a Mott insulator}, arXiv:1904.10995 (2019).

\bibitem{NoteRatioVolume} The volume occupied by the scattering sphere is $V_{\rm scatt} \simeq 4 \pi (k_{a}/2)^2 \times \delta  k_{s} $ (a sphere of radius $k_{a}/2$ and width $\delta k_{s}$), while the volume occupied by a peak of the condensate is $V_{\rm peak} \simeq 4 \pi \Delta k^3/3$ (a sphere of radius $\Delta k$). In the regime where the mean-field interaction are negligible, one expects $\delta k_{s}=\Delta k$. The ratio $V_{\rm scatt}/V_{\rm peak}$ is then $\sim(k_{a}/ \Delta k)^2 = (L/a)^2\gg1$ where $L/a$ is the number of lattice sites occupied by the trapped BEC. 

\bibitem{chang2016} R. Chang, Q. Bouton, H. Cayla, C. Qu , A. Aspect, C. I. Westbrook and D. Cl\'ement, {\it Momentum-resolved observation of thermal and quantum depletion in an interacting Bose gas}, Phys. Rev. Lett. {\bf 117}, 235303 (2016).

\bibitem{band2000} Y. B. Band, Marek Trippenbach, J. P. Burke, Jr., and P. S. Julienne, {\it Elastic Scattering Loss of Atoms from Colliding Bose-Einstein Condensate Wave Packets}, Phys. Rev. Lett. {\bf 84}, 5462 (2000).

\bibitem{zin2005} P. Zin, J. Chwedenczuk, A. Veitia, K. Rzazewski, and M. Trippenbach, {\it Quantum Multimode Model of Elastic Scattering from Bose-Einstein Condensates}, Phys. Rev. Lett. {\bf 94}, 200401 (2005).

\bibitem{zin2006} P. Zin, J. Chwedenczuk, and M. Trippenbach, {\it Elastic scattering losses from colliding Bose-Einstein condensates}, Phys. Rev. A {\bf 73}, 033602 (2006).

\bibitem{kramer2002} M. Kramer, L. P. Pitaevskii and S. Stringari, {\it Macroscopic dynamics of a trapped Bose-Einstein condensate in the presence of 1D and 2D optical lattices}, Phys. Rev. Lett. {\bf 88}, 180404 (2002).

\bibitem{stenger1999} J. Stenger, S. Inouye, A. P. Chikkatur, D. M. Stamper-Kurn, D. E. Pritchard, and W. Ketterle, {\it Bragg Spectroscopy of a Bose-Einstein Condensate}, Phys. Rev. Lett. {\bf 82}, 4569 (1999).

\bibitem{ashcroft1976} W. N. Ashcroft and N. Mermin, {\it Solid-state physics}, Brooks/Cole Publishing Company (1976).

\bibitem{NoteScattHalo} To calculate the number of collisions $N_{coll}^{\rm exp}$, we use only 5 of the 6 scattering halos as one of them is partially falling outside the physical surface of the He$^*$ detector.

\end{thebibliography}

%%%%%%%%%%%%%%%%%%%%%%

\end{document}